\documentclass[lettersize,journal]{IEEEtran}
\usepackage{amsmath,amsfonts}
\usepackage{algorithmic}
\usepackage{cite}
\usepackage{array}
\usepackage[caption=false,font=normalsize,labelfont=sf,textfont=sf]{subfig}
\usepackage{textcomp}
\usepackage{stfloats}
\usepackage{url}
\usepackage{verbatim}
\usepackage{graphicx}
\usepackage{xcolor}
\usepackage{multirow}
\def\BibTeX{{\rm B\kern-.05em{\sc i\kern-.025em b}\kern-.08em
		T\kern-.1667em\lower.7ex\hbox{E}\kern-.125emX}}
\usepackage{balance}
\begin{document}
	\title{\huge{Intelligent Reflecting Surface-Enabled Anti-Detection for Secure Sensing and Communications}}
\author{ 
	Beixiong Zheng,~\IEEEmembership{Senior Member,~IEEE}, Xue Xiong, Tiantian Ma,
	Jie Tang,~\IEEEmembership{Senior Member,~IEEE},\\
	Derrick Wing Kwan Ng,~\IEEEmembership{Fellow, IEEE}, 
	A. Lee Swindlehurst,~\IEEEmembership{Fellow,~IEEE}, 
	and Rui Zhang,~\IEEEmembership{Fellow,~IEEE}
\thanks{
B. Zheng, X. Xiong, T. Ma, and J. Tang are with South China University of Technology; 
D. W. K. Ng is with University of New South Wales;
A. L. Swindlehurst is with University of California, Irvine;
R.~Zhang is with Chinese University of Hong Kong, Shenzhen.
} 
}
	
	
	\maketitle 
	
\begin{abstract}
The ever-increasing reliance on wireless communication and sensing has led to growing concerns over the vulnerability of sensitive information to unauthorized detection and interception. 
Traditional anti-detection methods are often inadequate, suffering from limited adaptability and diminished effectiveness against advanced detection technologies. To overcome these challenges, this article presents the  intelligent reflecting surface (IRS) as a groundbreaking technology for enabling flexible electromagnetic manipulation, which has the potential to revolutionize anti-detection in both electromagnetic stealth/spoofing (evading radar detection) and covert communications (facilitating secure information exchange). We explore the fundamental principles of IRS and its advantages over traditional anti-detection techniques and discuss various design challenges associated with implementing IRS-based anti-detection systems. Through the examination of case studies and future research directions, we provide a comprehensive overview of the potential of IRS technology to serve as a formidable shield in the modern wireless landscape.
\end{abstract}
	
%

\section{Introduction}
Security remains a paramount concern in wireless systems, especially as future sixth-generation (6G) wireless networks evolve to provide ubiquitous sensing and massive communication capabilities \cite{Nguyen2021Security}. 
The inherent broadcast nature of the wireless medium, coupled with advancements in detection technologies, exposes sensitive information to a broad range of security threats \cite{Nguyen2021Security,Ren2023Robust,wang2023applications}.
For instance, aircraft, vehicles, and vessels are subject to radar detection revealing their location, direction, speed, etc. Similarly, the wireless transmission of confidential information (e.g., trade secrets, personal identities, military intelligence) may be intercepted or accessed without permission.
This potential exposure grows increasingly alarming as our reliance on sensitive information deepens across diverse applications in 6G, including commercial and military communication/sensing systems.
In certain practical scenarios, the failure to maintain confidentiality not only poses the risk of significant economic losses but also potentially results in setbacks on the battlefield, thereby jeopardizing national security.

In light of the above concerns, substantial efforts have been dedicated towards bolstering secure sensing and communications, with a specific focus on two prominent techniques: electromagnetic stealth/spoofing and covert communication \cite{ahmad2019stealth,fang2023diverse,du2022practical,yan2019low}.
In the context of secure sensing, electromagnetic stealth or spoofing attempts to evade radar detection by substantially reducing the target's radar cross section (RCS) and/or generating spoofing signals to mimic genuine targets to deceive radar systems \cite{ahmad2019stealth,fang2023diverse,du2022practical}.
However, traditional stealth and spoofing methods,  which rely on electromagnetic wave absorbing materials (EWAM) or sophisticated signal processing, inherently exhibit limited performance due to various practical implementation constraints and their inability to swiftly adapt to rapidly evolving environments. 
As for secure communication, traditional covert techniques leverage the inherent randomness of the wireless medium to conceal transmitted signals in environmental or artificial noise, thereby evading detection by malicious eavesdroppers \cite{yan2019low}. However, these techniques often involve high computational complexity or exhibit limited scalability across various practical deployment scenarios.
Furthermore, they primarily rely on predictable signal manipulation patterns that are susceptible to advanced eavesdropping techniques, e.g., machine learning-based schemes.
Consequently, traditional anti-detection approaches for both secure sensing and communications not only exhibit limited efficacy across both the spatial and temporal domains, but also struggle with adaptability and flexibility against fast-varying environments and advanced detection algorithms.

\begin{figure*}[!t]
	\centering
	\includegraphics[width=1\textwidth]{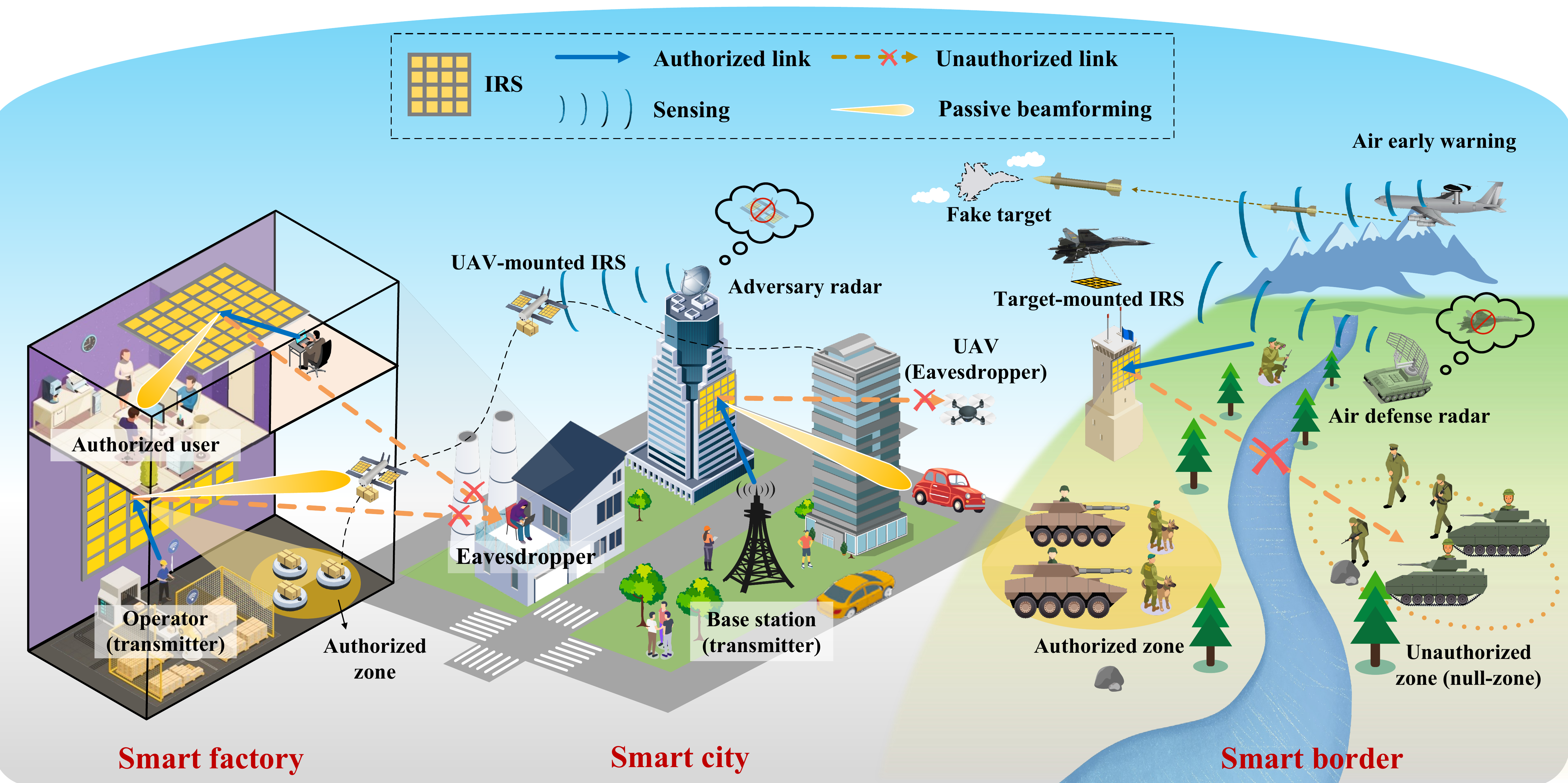}
	\caption{Typical scenarios for IRS-aided electromagnetic stealth/spoofing and covert communication systems.}
	\label{scenes}
\end{figure*}
Driven by recent advancements in digitally-controlled metasurfaces, intelligent reflecting surface (IRS) has emerged as a disruptive technology that enables smart and reconfigurable wireless environments in a cost-effective manner \cite{wu2021intelligent,zheng2022survey,wu2023intelligent}. An IRS essentially consists of a planar surface comprising numerous passive reflecting elements, each of which can be independently controlled to alter the amplitude and/or phase of the impinging signals. This capability allows for the elements' collaborative reshaping of the electromagnetic wave propagation in a customized manner.
Through fine-grained control of the surface reflection properties, IRS can effectively enhance the transmission according to actual system requirements and perform a variety of functions such as beam steering and interference cancellation, providing great potential for improving the performance and functionality of future anti-detection technologies.
In particular, IRS offers several advantages that can potentially enhance anti-detection capabilities: 
    \begin{itemize}
       	\item \textbf{Dynamic and adaptive control:} The IRS reflection patterns can be dynamically manipulated to deteriorate the received signal strength or introduce randomness for adversaries, thereby enabling adaptation to rapidly varying wireless environments \cite{wu2021intelligent}. 
    	\item \textbf{Multi-frequency and multi-angle operation:} By dynamically adjusting the reflection coefficients across multiple frequencies and directions, IRS can effectively nullify the received signals at various potential adversaries with distinct detection frequencies or location angles \cite{zheng2023intelligent}.
    	\item \textbf{Cost-effectiveness and scalability:} Since IRS is implemented using a fully passive surface with lightweight and conformal geometry \cite{zheng2022survey}, it provides benefits due to lower power consumption, reduced implementation complexity, and broader applicability compared to traditional anti-detection measures.
    	\item \textbf{Flexible integration:} Leveraging its exceptional compatibility with established wireless technologies \cite{wu2023intelligent}, IRS can serve as an auxiliary component and be seamlessly integrated into existing platforms without compromising performance.
    \end{itemize}
    
   In this context, IRS stands out as a competitive candidate technology to address the inherent challenges posed by traditional electromagnetic stealth/spoofing and covert communication techniques. This article aims to explore the potential of IRS-assisted electromagnetic stealth/spoofing and IRS-enabled covert communication techniques for enhancing secure sensing and communication, with an emphasis on their fundamental principles, design considerations, and potential challenges. 
   As depicted in Fig.~\ref{scenes}, we envision diverse applications of IRS-aided anti-detection technologies for improving the security performance of wireless systems across three typical scenarios, namely, smart factories, smart cities, and smart borders. 
   Moreover, this article provides representative case studies to further demonstrate the benefits of integrating IRS into secure sensing/communication architectures, and presents promising research directions to inspire future advancements in this emerging field.

    \renewcommand{\arraystretch}{1.5} 
\renewcommand{\arraystretch}{1.5} 
\begin{table*}[!t]
	\caption{Comparison of IRS-aided electromagnetic stealth and IRS-aided electromagnetic spoofing\label{tab:comp}}
	\centering
	\begin{tabular}{|m{3.3cm}<{\centering}|m{2.5cm}<{\centering}|m{2.5cm}<{\centering}|m{3cm}<{\centering}|m{2.4cm}<{\centering}|} 	
		\hline
		\bf Radar evasion technique &  \bf Purpose &  \bf Core idea & \bf Required information &  \bf IRS design strategy \\
		\hline \hline
		IRS-aided electromagnetic stealth \cite{zheng2023intelligent} & Make the target invisible to radars & Minimize the reflected signal power towards radars & Channels between target and radars (AoA, path gain) & Signal cancellation \\
		\hline
		IRS-aided electromagnetic spoofing & Deceive radars with false information about the target & Disguise reflected signals as echoes from a decoy target & Channels among target, radars, and scatterers (AoA, path gain) & Introduction of false information \\
		\hline
	\end{tabular}
\label{table_comp}
\end{table*}
    \section{IRS-Aided Electromagnetic Stealth/Spoofing}
\begin{figure}[!t]
	\centering
	\includegraphics[width=3.5in]{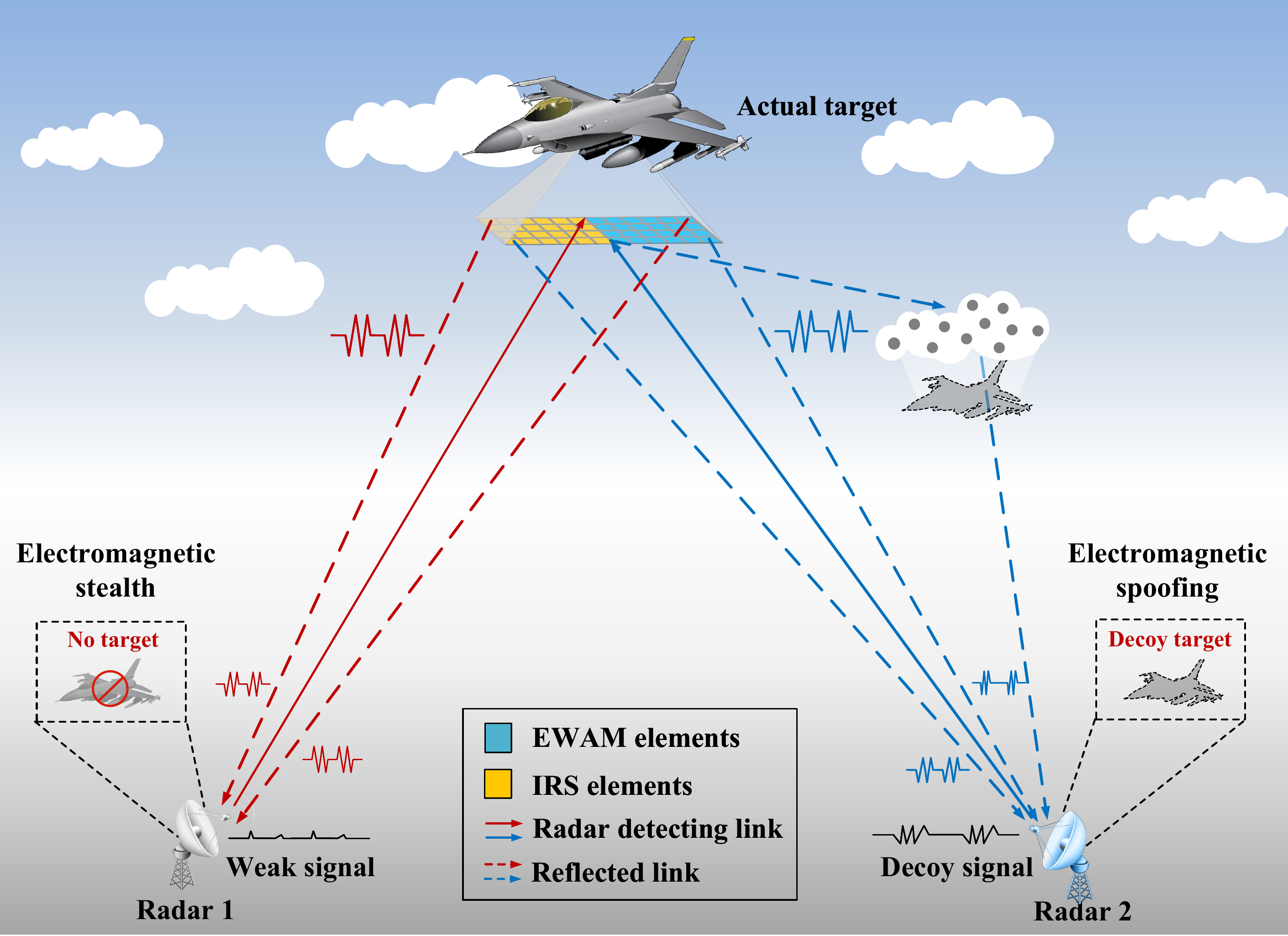}
	\caption{An IRS-aided electromagnetic stealth/spoofing system.}
	\label{ES}
\end{figure}

    Electromagnetic stealth \cite{ahmad2019stealth} and electromagnetic spoofing \cite{du2022practical} are two pivotal strategies in secure sensing aimed at anti-detection, sharing certain similarities yet possessing distinct characteristics as summarized in Table \ref{table_comp}. 
    Considering the limited flexibility and adaptability of traditional stealth/spoofing technologies, especially in rapidly evolving detection environments, IRS exhibits the potential to enable intelligent electromagnetic stealth/spoofing with real-time control across a wide range of operating frequencies and incident angles. 
    The fundamental concept behind IRS-aided stealth/spoofing is to manipulate the IRS reflection to neutralize or disguise the signals that are echoed back to adversarial radars, thereby diminishing the likelihood of their accurate detection \cite{zheng2023intelligent,shao2023target,Xiong2024ANew}.
    A typical scenario is shown in Fig.~\ref{ES}, where an IRS is mounted on a target surface to assist in evading/spoofing radar detection.
    By appropriately adapting the amplitude and/or phase shift at the IRS, the signals reflected by both the IRS elements and target surface can be destructively combined to cancel the signals echoed back to Radar 1, thereby effectively reducing the target's detectability (i.e, stealth). 
    Alternatively, the IRS can be configured to eliminate the signals reflected towards Radar~2 to shield the actual target, while simultaneously directing its reflected signals towards other surrounding scatterers to create decoy targets for Radar~2, thereby increasing the probability of false detection (i.e, spoofing).


In comparison to conventional stealth/spoofing technologies, IRS distinguishes itself through enabling flexibility, adaptability, and real-time reconfigurability against radar detection. One of the key advantages of IRS lies in its ability to dynamically adjust reflection in real-time, providing an effective solution for addressing the time-varying and highly dynamic detection environments due to high-mobility targets and the rapidly changing modes of adversarial radars \cite{zheng2023intelligent,shao2023target,Xiong2024ANew}. Additionally, IRS possesses the capability to manipulate its reflection patterns across a wide range of frequencies and incident angles employed by various radar systems \cite{zheng2023intelligent}. Moreover, benefiting from its low profile, light weight, and conformal geometry, IRS can be seamlessly integrated into existing stealth/spoofing platforms \cite{wu2023intelligent}. As a result, IRS can serve as an important component in developing intelligent and adaptive anti-detection strategies for secure sensing to cope with increasingly sophisticated radar detection scenarios.
    
Generally speaking, an IRS-aided stealth/spoofing system operates in two alternating modes: sensing mode for radar reconnaissance and reflection mode for electromagnetic stealth/spoofing. The sensing mode is first conducted to estimate radar information such as the angle-of-arrival (AoA) and path gain of the radar probing signals, which can be achieved leveraging dedicated sensors mounted on the target. 
Exploiting this estimated information, the IRS performs corresponding adjustments to its reflection coefficients to achieve stealth/spoofing.
For stealth systems, the IRS reflection can be customized with the aim of signal cancellation to maximize invisibility. In this context, the phase shifts of the reflected waves can be manipulated either continuously or discretely to effectively nullify the target's reflected signal received by the radar. 
When dealing with multiple distributed radars, IRS reflection design  needs to take into account the combined effect of all incoming radar signals arriving from different directions in general. This can be addressed with an angle-selective reflection pattern, which simultaneously suppresses multiple probing signals echoed back to different angles.
On the other hand, when dealing with uncertain adversarial radar locations within a specific region, adaptive null steering becomes an effective reflection strategy for concealing the target. 
This strategy involves adapting the reflection pattern to forge ``null" zones with minimal signal strength, which effectively act like a dynamic ``spatial filter," shielding specific directions from the reflected signals.
For spoofing systems, building upon the aforementioned stealth strategies to protect the actual target, the IRS can also redirect its reflected signals towards surrounding scatterers to create decoy targets to confuse radar detection.
The IRS reflection pattern can be modulated or randomized to introduce additional deceptive information into  the reflected signals, thereby generating false target echoes to deceive adversarial radar systems. 

\section{IRS-Enabled Covert Communications}
\begin{figure}[!t]
	\centering
	\includegraphics[width=3.6in]{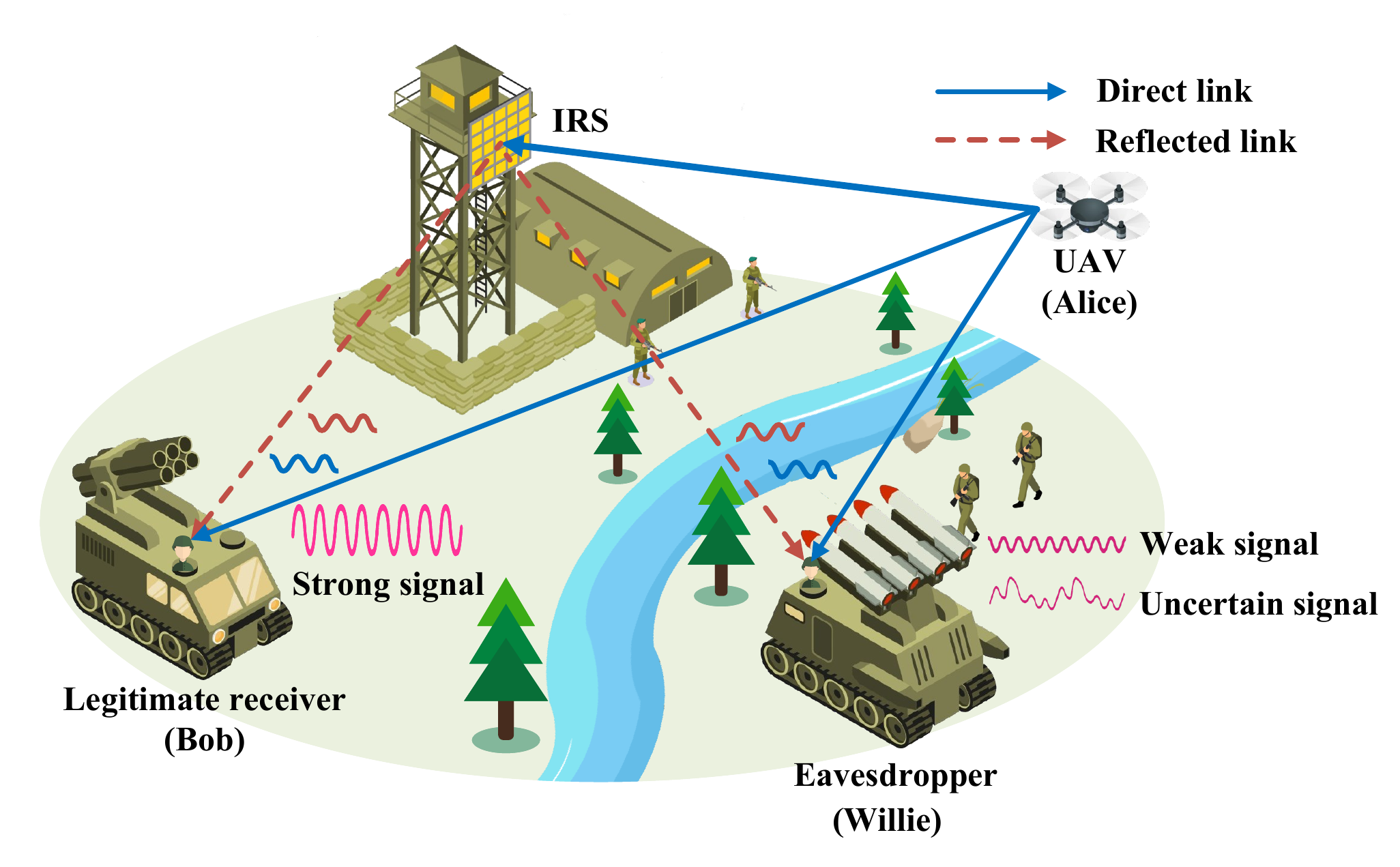}
	\caption{An IRS-enabled wireless covert communication system.}
	\label{Cov_comm}
\end{figure} 
Electromagnetic stealth/spoofing and covert communication share a common design objective: enhancing security by evading/deceiving adversary detection. However, they achieve this goal through distinct approaches.
Specifically, stealth/spoofing techniques are designed to prevent radar detection by reducing the target's RCS and/or generating confusing echoed signals, while covert communications seek to hide the existence of wireless transmission by introducing uncertainties, such as random noise and channel fluctuations, for unauthorized users \cite{yan2019low}. 
In practice, traditional covert technologies, such as artificial noise generation, directional antennas, and spread spectrum, inevitably consume additional resources, thereby increasing the burden on the system design.
Fortunately, IRS offers a cost-effective alternative for enhancing covert communication performance. 
Employing a similar rationale to that in stealth/spoofing systems, the core idea of IRS-enabled covert communication is to intelligently manipulate wireless propagation environments around the legitimate and eavesdropping users, rendering sensitive communication signals undetectable to unauthorized receivers \cite{zhou2021intelligent,lu2020intelligent}.
As shown in Fig.~\ref{Cov_comm}, in a typical IRS-enabled covert communication system, the transmitter (Alice) intends to covertly transmit information to the legitimate receiver (Bob) with the aid of an IRS, while an unauthorized receiver (Willie) aims to detect the existence of this transmission. 
By strategically calibrating the reflection coefficients of the IRS, the signal reflected by the IRS and the direct signals transmitted from Alice can not only be constructively combined at Bob to enhance the legitimate transmission rate, but also be destructively superimposed at Willie to prevent potential information leakage.
In addition, the IRS phase shifts can be designed to deliberately introduce uncertainties into the eavesdropping channels, thereby increasing Willie's probability of detection error.

Compared to traditional covert communication techniques that rely on predictable patterns of signal manipulation, IRS provides several compelling advantages.  
First, IRS offers fine-grained signal control, effectively enhancing the signal strength of covert transmission channels while simultaneously reducing signal strength or introducing uncertainty to obfuscate unauthorized entities.
Second, IRS enables adaptive and reconfigurable covert communication by dynamically adjusting its reflection pattern in response to instantaneous channel conditions, user positions, and evolving eavesdropping behaviors.
Third, IRS can lead to a higher covert transmission rate by manipulating signals across a broader bandwidth and effectively concealing information over different frequencies to prevent unauthorized receivers from accessing sensitive information.
Finally, IRS can enhance covert robustness by generating a ``signal shield" around the legitimate communication zone that significantly attenuates radiated signals and reduces information leakage to the unauthorized zone.
Overall, integrating IRS into covert communication systems greatly enhances flexibility, real-time adaptability, and the ability to shield specific communication regions, thereby improving covert communication performance.
	
In IRS-enabled covert communication systems, the knowledge available to the legitimate transmitter, Alice, regarding the eavesdropper's channel state information (CSI) plays a crucial role in determining the optimal IRS reflection pattern. 
As such, a reconnaissance mode is initially conducted at the transmitter and/or sensing devices installed on the IRS to acquire the cascaded or separate CSI of both the legitimate and eavesdropping users. 
With the estimated CSI, the reflection coefficients are then optimized to facilitate the optimal design of the passive beamforming, enabling the concentration of transmitted signals toward the legitimate user for enhancing achievable rates, while simultaneously nulling them toward potential eavesdroppers to minimize detection probability. 
Alternatively, one can introduce uncertainty in the IRS phase shifts to mask the origin and content of the communication signals, thus leading to increased probability of erroneous detection by unauthorized receivers. 
Moreover, IRS deployment will directly impact the covert transmission mechanism and system coverage. 
For example, centralized IRS deployment strategies may provide a higher transmission rate by leveraging a higher passive beamforming gain for the legitimate receiver, while distributed IRS deployment strategies lead to wider coverage and higher flexibility for covert communications. 
In addition, different IRS deployment strategies generally impose distinct requirements on IRS channel estimation and reflection design, which necessitates customized deployment strategies under different system setups. 
	
\section{Design Challenges: Unlocking the Full Potential of IRS for Anti-Detection}
Despite the appealing advantages and significant potential of IRS, several design issues and challenges need to be addressed for IRS-aided anti-detection to enable secure sensing and communication systems.
In the following, we elaborate on the main challenges associated with IRS reflection design, reconnaissance sensing, as well as IRS deployment, while also providing forward-looking solutions and directions for future exploration.

\subsection{IRS Reflection Design}
To establish high-quality IRS-enabled signal shielding for secure sensing and communications, the IRS reflection coefficients need to be carefully designed to minimize the probability of intercept/detection by adversaries (e.g., radars and eavesdroppers) \cite{zheng2023intelligent,shao2023target,Xiong2024ANew,zhou2021intelligent,lu2020intelligent}. 
The integration of IRS with anti-detection technologies brings new paradigms for secure sensing and communications, resulting in novel optimization problems for IRS reflection design. 
These optimization problems are generally non-convex and challenging to be solved optimally due to the constant-modulus constraints associated with the large number of IRS reflecting elements. 
A straightforward approach to this issue is to employ semidefinite relaxation (SDR) techniques for relaxing the non-convex constant-modulus constraints; however, this may incur excessively high computational complexity \cite{Xiong2024ANew}. 
On the other hand, to address the coupling between the reflection and transmit beamforming, various efficient approaches, such as alternating optimization (AO) and block coordinate descent (BCD) algorithms, can be employed to obtain suboptimal IRS solutions with polynomial-time complexity \cite{wu2021intelligent}. 
Furthermore, the acquisition of IRS-related CSI of radars/eavesdroppers may be inaccurate or even unavailable in practice, which can degrade the performance of IRS-aided anti-detection techniques.
This presents a formidable challenge in ensuring reliable stealth/spoofing and covertness performance amidst various imperfections, especially when employing conventional reflection strategies assuming the availability of perfect CSI. 
In this context, it is necessary to investigate robust IRS reflection design under different levels of CSI availability/uncertainty. 

Another design challenge arises when the exact locations of adversaries in unauthorized areas are unknown, which complicates the IRS reflection design.
 A potential solution to this issue is by establishing a shielded zone around such areas where the reflected signals are effectively attenuated or redirected away.
In addition, one can introduce randomized reflection patterns to maximize the false detection probability of adversaries within this designated area. 
Dealing with high-mobility targets or users also presents a unique challenge. 
In particular, the time available to update the IRS reflection pattern is limited, making real-time IRS reflection design quite challenging. 
To mitigate this issue, one possible solution is to calculate the IRS reflection coefficients offline and store them in the IRS controller. 
With this precalculated database, the IRS can dynamically adjust its signal reflection in real-time in response to the sensed AoA information. 
Artificial intelligence (AI) algorithms, such as federated and reinforcement learning, can be employed to learn offline the approximate mapping from the AoA information to the optimal IRS reflection coefficients.  
Another promising approach is to leverage adaptive methods with low computational complexity, such as the Kalman filter, for timely updates of the IRS reflection coefficients over time. 
Given the aforementioned challenges, there is a need for deeper investigation into highly efficient IRS reflection designs in future research endeavors.
	
\subsection{Sensing for Reconnaissance}
Acquiring accurate CSI is indispensable for ensuring the appropriate and real-time adjustment of the IRS reflection in response to environmental variations. However, CSI estimation is challenging in practice due to the massive number of passive reflecting elements without active components for baseband signal processing. To tackle this issue, the proposed electromagnetic stealth/spoofing system incorporates a sensing array embedded on a target-mounted IRS for radar reconnaissance. 
This setup facilitates the collection of environmental data, such as the number of radars as well as the AoA and path gain information for the radars and surrounding scatterers. The AoA information can be obtained by adopting well-established AoA estimation algorithms such as the multiple signal classification (MUSIC), estimation of signal parameters via rotational invariance  techniques (ESPRIT), and compressed sensing (CS) methods. Additionally, the path gains can be further estimated leveraging the maximum likelihood (ML) or least square (LS) estimation methods.
In covert communication systems, the CSI of eavesdroppers may not be easy to obtain if they intentionally remain silent during channel estimation.  
Moreover, when performing the IRS channel estimation for legitimate users, eavesdroppers could secretly learn or intercept the legitimate CSI and/or launch pilot spoofing or contamination attacks to impair the channel estimation for legitimate users.
To mitigate the risk of information leakage, one can conduct reconnaissance at the transmitter to monitor the eavesdroppers' behavior and simultaneously estimate the cascaded CSI of both the legitimate and eavesdropping users. 
Alternatively, reconnaissance can be conducted via IRS-mounted sensing devices to estimate the IRS-involved channels between the transmitter and legitimate/eavesdropping users, using similar estimation procedures as those exploited in stealth/spoofing systems.

On the other hand, the high mobility of targets/users gives rise to rapidly varying wireless channels as well as Doppler effects, which requires frequent channel estimation. This makes the CSI acquisition problem more challenging in IRS-aided secure sensing and communication systems. 
To mitigate this problem, CSI can be updated block-by-block by adopting efficient signal processing methods such as compressed sensing and deep learning within blocks of sufficiently short duration.  
However, the risk persists as CSI may become quickly outdated before the next block is received. 
Therefore, based on prior knowledge of the movement trajectory, real-time channel prediction/tracking methods could be employed to acquire future CSI, which can be realized for example through the application of Kalman filter-based algorithms. 
Nevertheless, these channel tracking methods may become inaccurate if the target/user's speed fluctuates dramatically or the estimation error and prediction error accumulate over time, which necessitates further CSI calibration. 
In a nutshell, these challenges underscore the importance of new research into cost-effective and high-resolution estimation/tracking methods for future advancements.


\subsection{Deployment of Sensors and IRSs}
The deployment of sensing devices and IRS elements significantly impacts the implementation complexity and cost of anti-detection technologies. Different sensor configurations, such as crossed-shaped, L-shaped, and (non)-uniform sensing arrays, offer varying levels of sensing resolution to acquire the CSI \cite{zheng2022survey}. Generally, increasing the number of sensors in a specific array improves the sensing resolution, but this comes at the expense of increased implementation cost and signal processing complexity. As such, a key deployment challenge lies in determining the optimal sensor configuration that can satisfy specific sensing requirements across various detection scenarios.

In addition, the deployment of IRS elements also significantly affects the cost-effectiveness, adaptability, and coverage range of the secure system. Thus, striking a balance among these factors is a practical challenge when designing IRS deployment strategies.
In stealth/spoofing systems, a primary deployment challenge lies in determining the optimal physical location of the IRS to achieve maximum effectiveness while minimizing cost.
Whether the target is stationary or moving, it is preferable to install IRS on the target surface, which enables adaptive cancellation/redirection of reflected radar signals from all directions by tuning the IRS reflection. 
In covert communication systems, one straightforward approach is to deploy IRS in a location that establishes strong line-of-sight (LoS) paths between the transmitter and legitimate receivers.
However, fixed IRS installations lack adaptability for mobile users. To overcome this limitation, a viable solution is to employ aerial IRS, for example on unmanned aerial vehicles (UAVs), that can be dynamically repositioned as required. This approach enables flexible coverage through dynamic passive beamforming, albeit necessitating additional costly infrastructure and power sources.
Another crucial deployment problem emerges from the limited coverage range offered by a single IRS or multiple co-located IRSs, which may fail to satisfy the quality of service (QoS) requirements of secure systems in complicated environments with many closely-spaced obstacles. 
Strategically placing multiple distributed IRSs allows for general multi-IRS reflections, thus providing more degrees of freedom (DoF) to sense adversaries and control the IRS reflection patterns from different perspectives. This also enables reaping more cooperative gain among IRSs to effectively deal with diverse adversaries.
Overall, designing well-informed deployment strategies based on specific threat scenarios and operational needs can yield optimal coverage performance in IRS-aided anti-detection wireless networks, opening up new avenues for future research dedicated to enhancing the effectiveness of secure systems in challenging environments.
	
\section{Case Studies}
\begin{figure}[!t]
	\centering
	\includegraphics[width=3.2in]{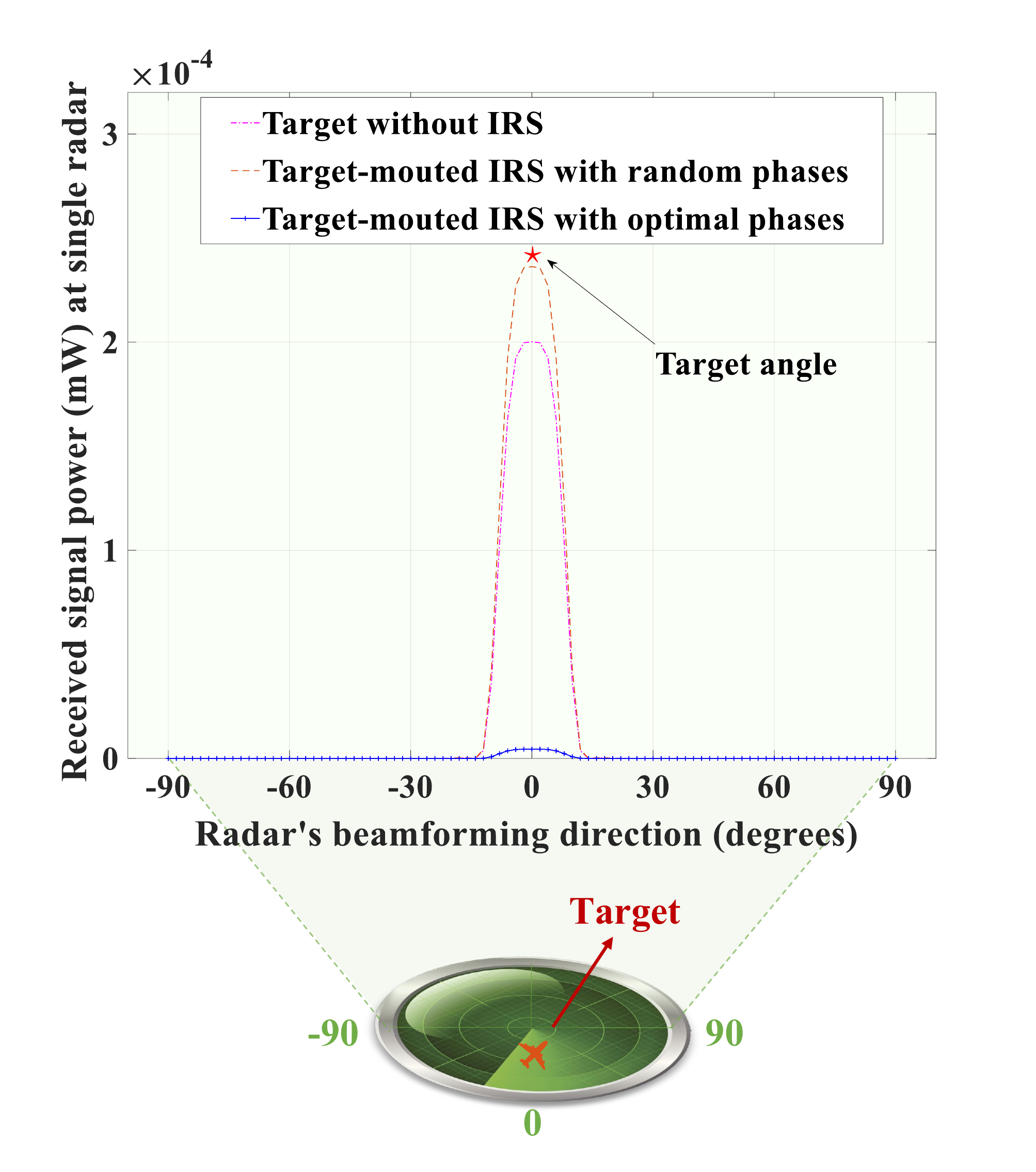}
	\caption{Performance of an IRS-aided electromagnetic stealth system against single-radar detection.}
	\label{single}
\end{figure}
To demonstrate the effectiveness of the proposed IRS-enabled anti-detection strategies, we consider two representative case studies for IRS-aided electromagnetic stealth and discuss their extension to other typical secure sensing and communication systems. 
We consider the system shown in Fig.~\ref{ES}, where an IRS with $N$ elements is mounted on a moving airborne target. For ease of exposition, we assume that perfect CSI is available to the IRS.
Meanwhile, the EWAMs coated on the target surface are modeled as a uniform planar array with 200 passive elements, each with an absorbing efficiency of $0.8$. 
Additionally, we assume that the adversarial radar system operates at 6 GHz, with transmit power $P = 15$ dBm, and each mono-static radar is equipped with $M=64$ transmit/receive antennas. 
Furthermore, to better illustrate the superior performance of the IRS-aided electromagnetic stealth system with the optimized reflection design as in \cite{zheng2023intelligent}, we compare it against two baseline systems: 1) Target without IRS and 2) Target-mounted IRS with random phase shifts.

\subsection{Single-Radar Case}

 \begin{figure}[!t]
	\centering
	\includegraphics[width=3.2in]{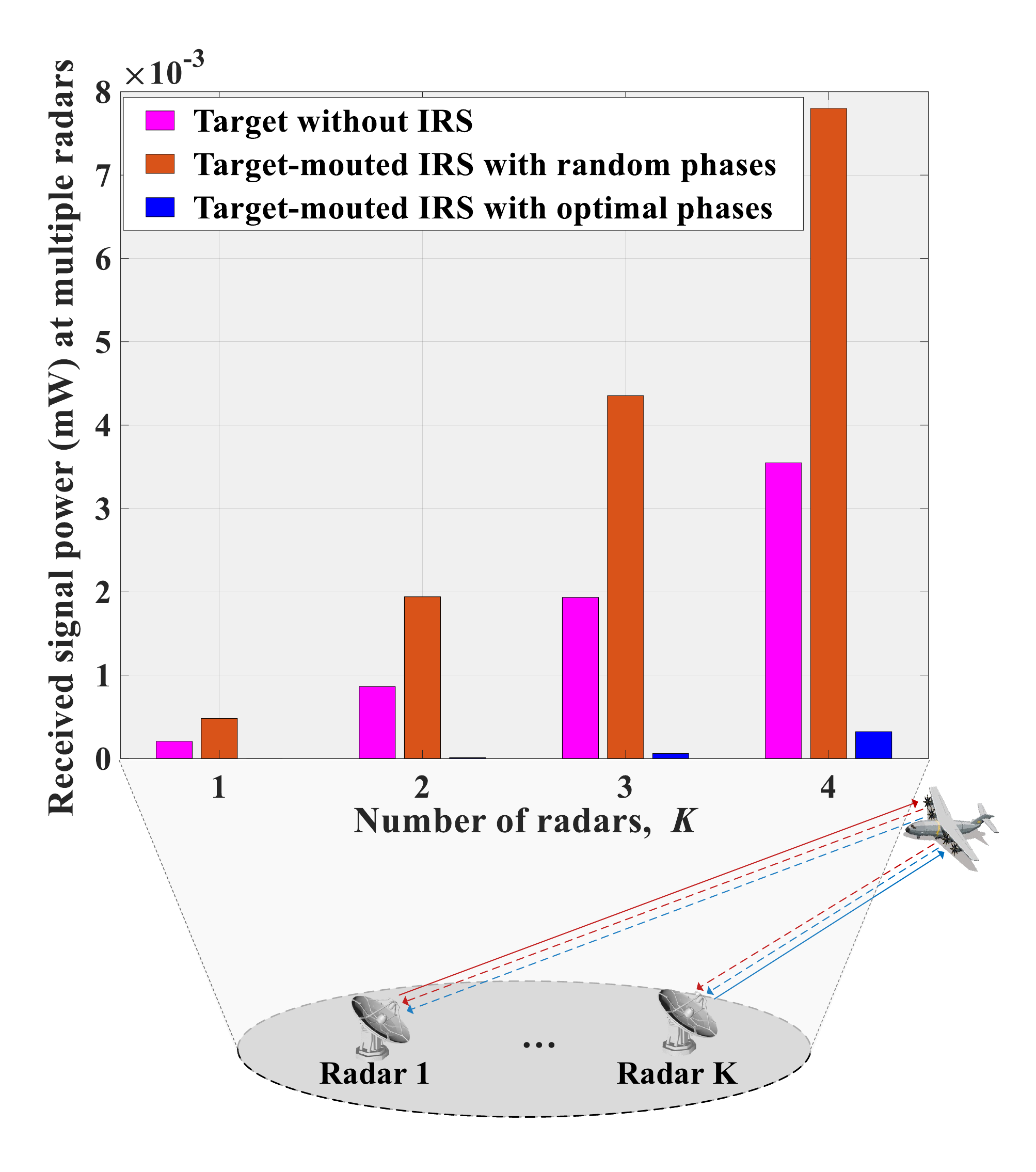}
	\caption{Performance of an IRS-aided electromagnetic stealth system against multi-radar detection.}
	\label{multiple}
\end{figure}
We first consider the case of detection with a single radar, where the target is situated directly above the radar and the number of IRS elements is set to $N=8$. We adopt the reverse alignment-based reflection design proposed in \cite{zheng2023intelligent} to achieve optimal stealth performance with low complexity.
As depicted in Fig.~\ref{single}, at the target angle $\vartheta =0^\circ$, the radar experiences a significant decrease in the received signal power due to the implementation of the optimized reflection design on the target-mounted IRS. In contrast, the received signal power remains much higher for the two baseline approaches.
This result verifies the effectiveness of incorporating an IRS with CSI-based reflection design in achieving electromagnetic stealth and suppressing the reflected signal to the single radar.
Moreover, this result can easily be extended to the IRS-aided covert communication system shown in Fig.~\ref{Cov_comm}, where an unauthorized receiver seeks to detect the existence of information exchange between a transmitter and receiver. By carefully designing the IRS reflection coefficients, the received signals at the unauthorized receiver can be deteriorated, thus cancelling the transmitted signals at the unauthorized receiver.
	
\subsection{Multi-Radar Case}


Next, we study a typical multi-radar scenario to evaluate the target-mounted IRS-enabled electromagnetic stealth system, as illustrated in Fig.~\ref{multiple}, where the number of IRS elements is $N=50$.
To handle the collective impact of all incoming signals from multiple radars, we utilize the minimum mean-square error (MMSE)-based design proposed in \cite{zheng2023intelligent} to obtain the near-optimal phases at the target-mounted IRS.
It is observed that as the number of radars increases, all systems exhibit a rise in the sum received signal power. 
This result is expected, since more distributed radars inevitably lead to an expanded collective reception capacity from multiple directions, which is manifested in the increase of their overall received power. 
Despite this, our proposed IRS-aided electromagnetic stealth system still demonstrates a significantly lower rate of increase in the sum received signal power compared to the two baseline systems without IRS and with random IRS phase shifts, respectively.
This result validates the ability of the IRS reflection design in simultaneously canceling the reflected signals across different directions to multiple radars.
This result can be readily extended to IRS-aided covert communication systems in the presence of multiple adversaries. By adjusting the IRS reflection beamforming, the signals transmitted from Alice to multiple directions can be attenuated or eliminated at different eavesdroppers' locations.
	
\section{Conclusions}
In this article, we provided a comprehensive exploration of IRS-enabled anti-detection technologies, specifically tailored for secure sensing and communications.
We first introduced the exploitation of IRS for electromagnetic stealth/spoofing and covert communication systems, shedding light on their fundamental principles and highlighting their advantages over traditional techniques. 
Subsequently, we examined the primary design challenges associated with harnessing IRS technology in secure sensing and communications, offering promising solutions as potential future directions. 
Finally, two case studies were presented to illustrate the effectiveness of integrating IRS into anti-detection systems. 
It is foreseen that IRS-enabled anti-detection will provide new research avenues in both secure sensing (electromagnetic stealth/spoofing) and covert communication in future 6G wireless networks.

\bibliographystyle{IEEEtran}
\bibliography{IRS_Stealth}

\begin{IEEEbiographynophoto}{Beixiong Zheng}
	 (bxzheng@scut.edu.cn) is an Associate Professor with the School of Microelectronics, South China University of Technology, Guangzhou, China.
\end{IEEEbiographynophoto}

\begin{IEEEbiographynophoto}{Xue Xiong}
	(ftxuexiong@mail.scut.edu.cn) is with the School of Microelectronics, South China University of Technology, Guangzhou, China.
\end{IEEEbiographynophoto}

\begin{IEEEbiographynophoto}{Tiantian Ma}
	(mitiantianma@mail.scut.edu.cn) is with the School of Microelectronics, South China University of Technology, Guangzhou, China.
\end{IEEEbiographynophoto}

\begin{IEEEbiographynophoto}{Jie Tang}
	(eejtang@scut.edu.cn) is a Professor with the School of Electronic and Information Engineering, South China University of Technology, Guangzhou, China.
\end{IEEEbiographynophoto}

\begin{IEEEbiographynophoto}{Derrick Wing Kwan Ng}
	[F] (w.k.ng@unsw.edu.au) is a Scientia Associate Professor with the University of New South Wales, Sydney, Australia.
\end{IEEEbiographynophoto}

\begin{IEEEbiographynophoto}{A. Lee Swindlehurst}
	[F] (swindle@uci.edu) is a Professor with the Department of Electrical Engineering and Computer Science, University of California, Irvine, USA.
\end{IEEEbiographynophoto}

\begin{IEEEbiographynophoto}{Rui Zhang}
	[F] (rzhang@cuhk.edu.cn) is a Chair Professor with The Chinese University of Hong Kong, Shenzhen, China.

\end{IEEEbiographynophoto}

\end{document}